\documentclass[fleqn,10pt]{wlscirep}
\usepackage[utf8]{inputenc}
\usepackage[T1]{fontenc}
\usepackage{multirow}

\usepackage{graphicx}
\usepackage{amsmath}
\usepackage{amssymb}
\usepackage{booktabs}
\usepackage{amssymb}
\usepackage{pifont}
\newcommand{\cmark}{\ding{51}}%
\newcommand{\xmark}{\ding{55}}%
\usepackage{makecell}
\usepackage[normalem]{ulem}

\definecolor{RevisionColor}{rgb}{1.0, 0.57, 0.0}

\title{A large-scale heterogeneous 3D magnetic resonance brain imaging dataset for self-supervised learning}

\author[1,2,3,+,*]{Stefano Cerri}
\author[1,2,+,*]{Asbj{\o}rn Munk}
\author[1,2]{Sebastian N{\o}rgaard Llambias}
\author[1,2]{Jakob Ambsdorf}
\author[1,2]{Julia Machnio}
\author[3,8]{Vardan Nersesjan}
\author[7]{Christian Hedeager Krag}
\author[4,6]{Peirong Liu}
\author[1,2]{Pablo Rocamora García}
\author[1,2]{Mostafa Mehdipour Ghazi}
\author[7,9]{Mikael Boesen}
\author[3,10]{Michael Eriksen Benros}
\author[4,5]{Juan Eugenio Iglesias}
\author[1,2]{Mads Nielsen}

\affil[1]{Department of Computer Science, University of Copenhagen, Denmark}
\affil[2]{Pioneer Centre For AI, Denmark}
\affil[3]{Copenhagen Research Centre for Biological and Precision Psychiatry, Mental Health Centre Copenhagen, Copenhagen University Hospital, Denmark}
\affil[4]{Athinoula A. Martinos Center for Biomedical Imaging, Massachusetts General Hospital and Harvard Medical School, USA}
\affil[5]{Computer Science and Artificial Intelligence Laboratory, Massachusetts Institute of Technology, USA}
\affil[6]{Johns Hopkins University, USA}
\affil[7]{Radiological AI Testcenter, Denmark}
\affil[8]{Copenhagen University Hospital, Rigshospitalet, Denmark}
\affil[9]{Copenhagen University Hospital, Bispebjerg \& Frederiksberg Hospital, Denmark}
\affil[10]{Department of Clinical Medicine, Faculty of Health and Medical Sciences, University of Copenhagen, Denmark}

\affil[+]{\{stce,asmu\}@di.ku.dk}

\affil[*]{Equal contribution. Author order may be adjusted for individual use.}

\begin{abstract}

We present FOMO260K, a large-scale, heterogeneous dataset of 260,927 brain Magnetic Resonance Imaging (MRI) scans from 77,589 MRI sessions and 55,378 subjects, aggregated from 910 publicly available sources. The dataset includes both clinical- and research-grade images, multiple MRI sequences, and a wide range of anatomical and pathological variability, including scans with large brain anomalies. Minimal preprocessing was applied to preserve the original image characteristics while reducing entry barriers for new users. Companion code for self-supervised pretraining and finetuning is provided, along with pretrained models. FOMO260K is intended to support the development and benchmarking of self-supervised learning methods in medical imaging at scale.

\end{abstract}
\begin{document}

\flushbottom
\maketitle

\thispagestyle{empty}

\section*{Background \& Summary}

Self-supervised learning (SSL) has led to major breakthroughs in computer vision and natural language processing, largely driven by the availability of large-scale public datasets such as ImageNet~\cite{Deng2009}, Places365~\cite{zhou2017places}, and OpenWebText~\cite{Gokaslan2019OpenWeb}. These resources have enabled the development, benchmarking, and rapid iteration of powerful SSL methods under standardized settings. In neuroimaging, however, the lack of comparably large and diverse public datasets has slowed the adoption and evaluation of SSL approaches. Existing large-scale datasets, such as ADNI~\cite{Mueller2005}, UK Biobank~\cite{Bycroft2018}, PPMI~\cite{Marek2018}, and ABCD~\cite{Casey2018}, while valuable, are often curated for specific diseases or patient populations, and typically follow homogeneous imaging protocols with limited pathological variability. Access is often restricted by formal applications, strict data use agreements, and institutional approvals, and data are commonly distributed in formats that require domain-specific preprocessing (e.g., conversion from DICOM to NIfTI format or handling 4D acquisitions). These challenges raise the barrier to entry and hinder the scalability of SSL pretraining.

\begin{table}[!ht]
\centering
\scriptsize
\setlength{\tabcolsep}{4.5pt}

\begin{tabular}{l c c c c c c c c c c c }
\toprule
Dataset &
Scans &
\makecell{Sessions} &
\makecell{Co-\\registered} &
\makecell{Disease\\Metadata} &
\makecell{Clinical-grade\\(scans)} &
\makecell{Control} &
\makecell{Brain Tumor} &
\makecell{Stroke} &
\makecell{Mental\\Disorders} &
\makecell{Dementia} &
\makecell{Neurological\\Disorders} \\
\midrule
OpenMind          
& 114K 
& 46K (35K)
& \xmark 
& \xmark 
& 8K (7\%)
& 30K (85\%)
& 0.2K (1\%)
& 2.5K (7\%)
& 1.2K (4\%)
& 0 (0\%)
& 0.7K (2\%)
\\
FOMO45K (ours) 
& 46K  
& 12K (8K) 
& \cmark 
& \cmark 
& 5K (11\%) 
& 2.7K (34\%) 
& 2.7K (34\%) 
& 2.5K (30\%) 
& 0 (0\%) 
& 0.3K (3\%)
& 0 (0\%)
\\
FOMO260K (ours) 
& 261K 
& 78K (64K) 
& \xmark 
& \cmark 
& 53K (20\%)   
& 41K (64\%)   
& 16K (24\%)    
& 2.6K (4\%)    
& 2.1K (3\%)  
& 0.3K (1\%)
& 0.9K (1\%)
\\
\bottomrule
\end{tabular}
\caption{\textbf{Comparison between OpenMind, FOMO45K and FOMO260K}. 
FOMO260K encompasses a broader range of disorders and a higher proportion of low-resolution scans, which are typical in clinical practice. In particular, FOMO260K contains considerably more tumors compared to OpenMind. We use slice thickness as a proxy for classifying a scan as ``clinical-grade'', and define a clinical-grade scan as having a slice thickness above 3mm. We report disorder groups by number of sessions, derived from the disease metadata as described in \hyperref[sec:methods]{Methods}. Numbers in parentheses indicate the number of sessions with available disease metadata, and percentages are computed with respect to this subset. Note that nearly all scans in OpenMind and FOMO45K are included in FOMO260K.}
\label{tab:related_work}
\end{table}

To address these concerns, we introduce FOMO260K, a large-scale, heterogeneous dataset of brain MRI scans, comprising 260,927 scans from 77,589 sessions and 55,378 subjects, aggregated from 910 publicly available sources. FOMO260K includes both clinical- and research-grade imaging across multiple MRI sequences and captures a wide range of anatomical and pathological variability, including scans with large brain anomalies, making it more representative of real-world population-level data. 
FOMO260K builds on recent large-scale aggregation efforts such as OpenMind~\cite{Wald2025}, which provides an important and well-curated resource of 114K scans from OpenNeuro~\cite{OpenNeuro}. Almost 
all of these scans are also included in FOMO260K, which is more than two times larger and exhibits greater heterogeneity in imaging protocols and slice thickness, including a higher proportion of low-resolution scans typical of clinical practice. An overview of the differences between the two datasets is shown in Table~\ref{tab:related_work}.

FOMO260K is available as a single download from Hugging Face (\href{https://huggingface.co/datasets/FOMO-MRI/FOMO260K}{link}), together with code to preprocess the scans, incorporate additional datasets, perform self-supervised pretraining, and finetune models. This setup enables method development, standardized benchmarking, reproducible experiments, and broader adoption of SSL in medical imaging. An earlier release of this dataset, FOMO45K, was developed in parallel with the Foundation Model challenge at MICCAI 2025 (FOMO25)~\cite{Munk2026}, which aims to catalyze progress in self-supervised learning for medical imaging. Since then, the dataset has been significantly expanded to form FOMO260K, and it will continue to grow in future releases as additional cohorts and modalities become available.

\section*{Methods}
\phantomsection
\label{sec:methods}

\subsection*{Input data sources and acquisition}

\begin{figure}[t]
    \centering
    \includegraphics[width=\linewidth]{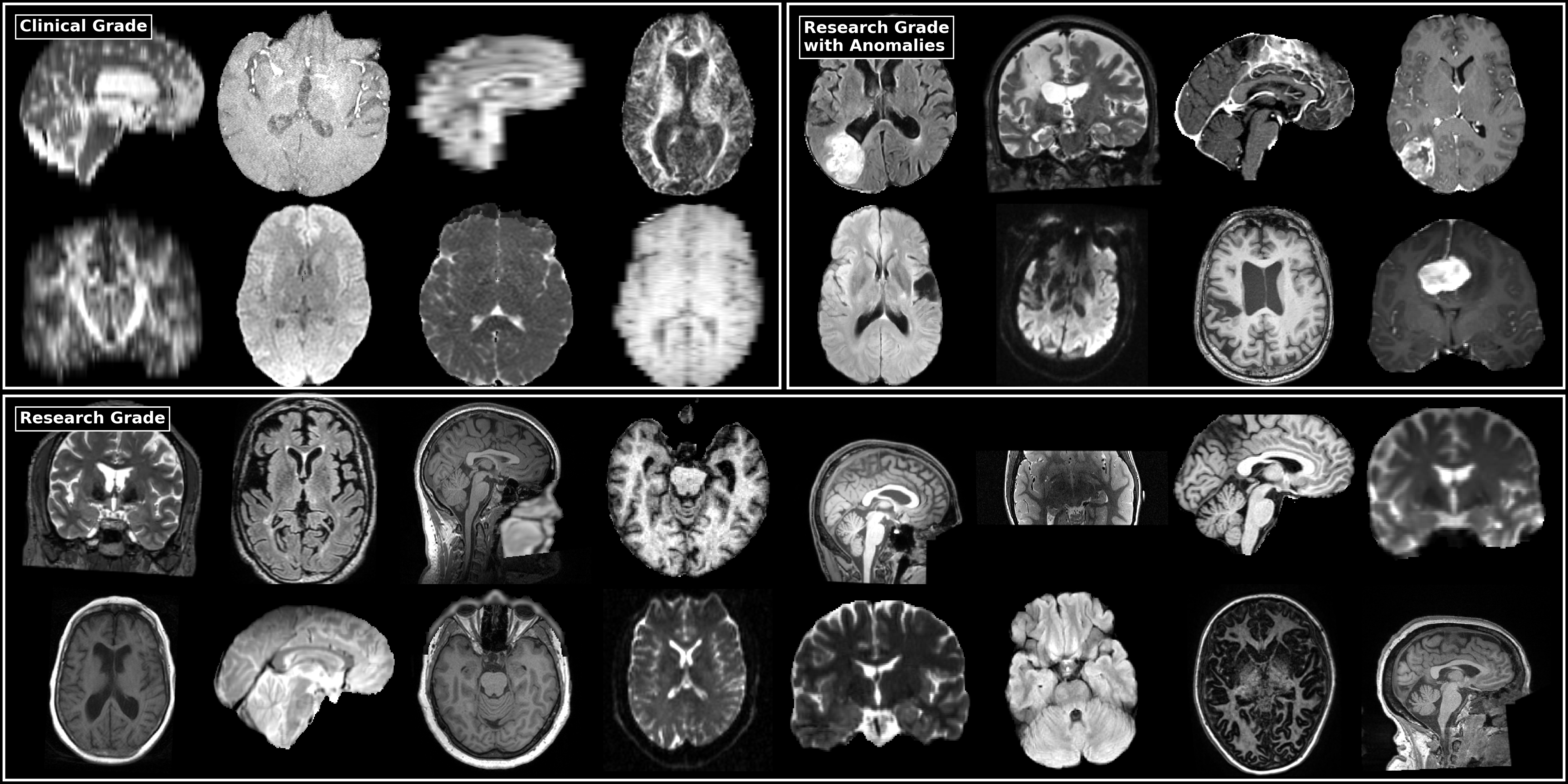}
    \caption{Representative examples from the FOMO260K dataset, illustrating the heterogeneity in image quality, MRI sequences, and the presence of brain anomalies.}
    \label{fig:fomo260kVisual}
\end{figure}

FOMO260K contains 260,927 scans from 77,589 MRI sessions and 55,378 subjects, aggregated from 910 publicly available sources. All available scans from each source were eligible for inclusion, except \textit{ex vivo} scans, functional MRI, field maps, SWI phase images, positron emission tomography scans, and computed tomography scans, which we excluded to prevent distribution shifts that could hinder SSL pretraining. No other exclusion criteria were applied. Table~\ref{tab:fomo260kDatasets} summarizes the source datasets, listing the number of subjects, sessions, scans, MRI sequence types, and license. All included datasets are publicly available under licenses compatible with CC BY-NC-SA redistribution, and the corresponding citations provide the DOI, URL, or accession identifier required to retrieve each original dataset. The largest contributions of FOMO260K come from HBN, Yale Brain Mets Longitudinal, and the OpenNeuro repository. T1-weighted structural MRI is present in 876 of the 910 constituent datasets (96.3\%), followed by T2-weighted (22.6\%), diffusion-weighted (13.2\%), FLAIR (3.5\%), and PD (2.2\%) sequences. A visual overview of the dataset’s heterogeneity---in terms of image quality, modality, and pathology---is provided in Figure~\ref{fig:fomo260kVisual}.

\begin{table}[!ht]
\centering
\scriptsize
\setlength{\tabcolsep}{6pt}
\begin{tabular}{l r r r l l}
\toprule
\makecell{\textbf{Source dataset}} &
\makecell{\textbf{Subjects}} &
\makecell{\textbf{MRI sessions}} &
\makecell{\textbf{MRI scans}} &
\makecell{\textbf{MRI sequences}} &
\makecell{\textbf{License}} \\
\midrule
    ClevelandCCF~\cite{ClevelandCCF, Biswal2010} & 31 & 31 & 31 & T1w & \href{https://fcon_1000.projects.nitrc.org/indi/retro/ClevelandCCF.html}{CC BY-NC} \\ 
    Nigerian Clinical~\cite{Wogu2025,Nigerian_clinical} & 82 & 140 & 701 & T1w, T2w, FLAIR, DWI & \href{https://brainlife.io/pub/675af884becd6464d43296c5}{CC BY 4.0} \\ 
    CUNMET~\cite{CUNMET} & 51 & 59 & 173 & T1w, M0, ASL & \href{http://fcon_1000.projects.nitrc.org/indi/retro/CUNMET.html}{CC BY-NC} \\ 
    ACPI~\cite{ACPI, Biswal2010} & 163 & 163 & 163 & T1w & \href{https://fcon_1000.projects.nitrc.org/indi/ACPI/html/acpi_mta_1.html}{CC BY-NC}\\ 
    ADHD\_200~\cite{ADHD200, Biswal2010} & 973 & 973 & 973 & T1w & \href{https://fcon_1000.projects.nitrc.org/indi/adhd200/}{CC BY-NC}\\ 
    AHEAD~\cite{Alkemade2020,Alkemade2020a} & 105 & 105 & 420 & T1w, T1map, R1map, R2*map & \href{https://uvaauas.figshare.com/articles/dataset/The_Amsterdam_Ultra-high_field_adult_lifespan_database_AHEAD_A_freely_available_multimodal_7_Tesla_submillimeter_magnetic_resonance_imaging_database/10007840}{CC BY 4.0} \\ 
    ATAG~\cite{Forstmann2014,Forstmann2014a} & 53 & 53 & 583 & T2*w, MP2RAGE & \href{https://datadryad.org/dataset/doi:10.5061/dryad.fb41s}{CC0} \\ 
    Adolescent Brain Development~\cite{Geeraert2020,Geeraert2020a} & 51 & 96 & 905 & T1w, T2w, DWI, ASL, CBF & \href{https://portal.conp.ca/dataset?id=projects/AdolescentBrainDevelopment}{PDDL}\\ 
    BraTS-GEN~\cite{BraTS-GEN, Labella2023, Adewole2023, baid2021, Moawad2024, Kazerooni2024} & 2,611 & 2,611 & 10,448 & T1w, T2w, FLAIR, T1ce & \href{https://www.synapse.org/Synapse:syn53708249/wiki/627759}{CC BY-NC} \\
    BrainLat~\cite{BrainLat, Prado2023} & 592 & 592 & 2,211 & T1w, FLAIR, DWI & \href{https://www.synapse.org/Synapse:syn51549340/wiki/624187}{CC0} \\ 
    CFMM-7T~\cite{Haast2021,Haast2021a} & 32 & 32 & 160 & T1w, MP2RAGE & \href{https://osf.io/k5zb9/}{CC BY 4.0} \\ 
    CHBMP~\cite{Valdes2021,Valdes2021a} & 203 & 203 & 901 & T1w, DWI & \href{https://portal.conp.ca/dataset?id=projects/CHBMP}{CC BY-NC-SA} \\ 
    Calgary Preschool~\cite{Reynolds2020,Reynolds2020a} & 162 & 473 & 1,629 & T1w, DWI, CBF & \href{https://osf.io/axz5r/}{CC BY 4.0} \\ 
    CoRR~\cite{CoRR} & 1,498 & 2,779 & 5,831 & T1w, MP2RAGE, DWI, ASL, CBF & \href{https://fcon_1000.projects.nitrc.org/indi/CoRR/html/concept.html}{Open Access}  \\ 
    MSD Brain Tumor~\cite{MSD, Simpson2019, Menze2014, Bakas2017, Bakas2018} & 750 & 750 & 3,000 & T1w, T2w, FLAIR, T1ce & \href{http://medicaldecathlon.com/}{CC BY-SA 4.0} \\ 
    HBN-SSI~\cite{HBN-SSI, Biswal2010}  & 13 & 168 & 1,031 & T1w, T2w, DWI & \href{https://fcon_1000.projects.nitrc.org/indi/cmi_healthy_brain_network/MRI_EEG.html#Direct%20Down}{CC BY-NC-SA 4.0}\\ 
    HBN~\cite{HBN, Alexander2017} & 3,847 & 3,849 & 26,097 & T1w, T2w, PD, FLAIR, DWI & \href{https://fcon_1000.projects.nitrc.org/indi/cmi_healthy_brain_network/MRI_EEG.html#Direct%20Down}{CC BY-NC-SA 4.0}\\ 
    IXI~\cite{IXI} & 584 & 584 & 3,105 & T1w, T2w, PD, DWI, ANGIO & \href{https://brain-development.org/ixi-dataset/}{CC BY-SA 3.0} \\ 
    Beijing Enhanced~\cite{BeijingEnhanced, Biswal2010} & 180 & 180 & 540 & T1w, DWI & \href{http://fcon_1000.projects.nitrc.org/indi/retro/BeijingEnhanced.html}{CC BY-NC}\\
    Infant Development Brain~\cite{Akinci2023, Akinci2023a} & 833 & 833 & 1,666 & T1w, T2w & \href{https://zenodo.org/records/8055666}{CC BY 4.0}\\ 
    M4Raw~\cite{Lyu2023,Lyu2023a} & 208 & 208 & 2,230 & T1w, T2w, FLAIR & \href{https://zenodo.org/records/8056074}{CC BY 4.0}\\ 
    MICA MICs~\cite{Royer2022,MICA-MICs} & 50 & 50 & 649 & T1w, T1map, MP2RAGE & \href{https://portal.conp.ca/dataset?id=projects/mica-mics}{CC0}\\ 
    NKI~\cite{Tobe2022,NKI} & 1,327 & 2,456 & 10,658 & T1w, T2w, DWI & \href{https://rocklandsample.org/for-researchers/step-2-data-usage-agreement}{Open Access}\\ 
    \multirow{4}{*}{OpenNeuro (884 datasets)~\cite{
                   ds000001,ds000002,ds000003,ds000005,ds000006,ds000007,
                   ds000008,ds000009,ds000011,ds000017,ds000030,ds000031,
                   ds000051,ds000052,ds000053,ds000101,ds000102,ds000105,
                   ds000107,ds000108,ds000109,ds000110,ds000113,ds000114,
                   ds000117,ds000119,ds000120,ds000121,ds000122,ds000133,
                   ds000140,ds000144,ds000148,ds000149,ds000157,ds000158,
                   ds000164,ds000168,ds000170,ds000171,ds000172,ds000174,
                   ds000200,ds000201,ds000202,ds000203,ds000204,ds000205,
                   ds000206,ds000208,ds000210,ds000212,ds000213,ds000214,
                   ds000216,ds000217,ds000218,ds000219,ds000220,ds000221,
                   ds000222,ds000223,ds000224,ds000228,ds000229,ds000231,
                   ds000232,ds000233,ds000234,ds000235,ds000236,ds000237,
                   ds000238,ds000239,ds000240,ds000241,ds000243,ds000244,
                   ds000245,ds000246,ds000247,ds000248,ds000249,ds000253,
                   ds000254,ds000255,ds000256,ds000258,ds001021,ds001037,
                   ds001110,ds001131,ds001132,ds001145,ds001168,
                   ds001178,ds001226,ds001228,ds001229,ds001232,
                   ds001233,ds001235,ds001241,ds001242,ds001246,ds001247,
                   ds001297,ds001299,ds001302,ds001306,ds001338,ds001339,
                   ds001344,ds001345,ds001353,ds001357,ds001365,ds001371,
                   ds001378,ds001379,ds001386,ds001399,ds001408,ds001415,
                   ds001417,ds001419,ds001420,ds001421,ds001430,ds001439,
                   ds001454,ds001486,ds001491,ds001497,ds001499,ds001506,
                   ds001510,ds001511,ds001517,ds001521,ds001525,ds001534,
                   ds001545,ds001547,ds001551,ds001553,ds001554,ds001555,
                   ds001563,ds001566,ds001576,ds001590,ds001595,ds001597,
                   ds001600,ds001607,ds001608,ds001612,ds001614,ds001621,
                   ds001634,ds001635,ds001652,ds001705,ds001715,ds001722,
                   ds001728,ds001734,ds001740,ds001743,ds001745,ds001747,
                   ds001748,ds001751,ds001761,ds001762,ds001771,ds001775,
                   ds001780,ds001784,ds001796,ds001814,ds001818,ds001832,
                   ds001838,ds001839,ds001840,ds001847,ds001848,ds001882,
                   ds001883,ds001894,ds001907,ds001912,ds001921,ds001923,
                   ds001926,ds001927,ds001928,ds001942,ds001946,ds001972,
                   ds001978,ds001984,ds002000,ds002001,ds002006,ds002011,
                   ds002013,ds002014,ds002033,ds002040,ds002041,ds002076,
                   ds002080,ds002105,ds002116,ds002149,ds002153,ds002155,
                   ds002156,ds002158,ds002168,ds002169,ds002185,ds002207,
                   ds002232,ds002236,ds002237,ds002241,ds002242,ds002250,
                   ds002270,ds002278,ds002293,ds002294,ds002295,ds002306,
                   ds002311,ds002316,ds002320,ds002322,ds002328,ds002330,
                   ds002336,ds002338,ds002345,ds002363,ds002366,ds002367,
                   ds002380,ds002382,ds002411,ds002419,ds002422,ds002424,
                   ds002425,ds002513,ds002522,ds002543,ds002547,ds002549,
                   ds002550,ds002574,ds002578,ds002596,ds002603,ds002606,
                   ds002608,ds002609,ds002614,ds002620,ds002634,ds002643,
                   ds002647,ds002655,ds002672,ds002674,ds002675,ds002684,
                   ds002685,ds002687,ds002702,ds002711,ds002715,
                   ds002717,ds002725,ds002726,ds002727,ds002731,ds002732,
                   ds002733,ds002734,ds002735,ds002737,ds002738,ds002739,
                   ds002741,ds002743,ds002748,ds002750,ds002766,ds002770,
                   ds002773,ds002776,ds002785,ds002790,ds002793,ds002797,
                   ds002799,ds002813,ds002814,ds002835,ds002837,ds002841,
                   ds002842,ds002843,ds002848,ds002872,ds002878,ds002879,
                   ds002886,ds002896,ds002898,ds002905,ds002936,ds002938,
                   ds002940,ds002979,ds002989,ds002994,ds002995,ds003007,
                   ds003011,ds003012,ds003017,ds003020,ds003037,ds003043,
                   ds003047,ds003059,ds003076,ds003078,ds003082,ds003083,
                   ds003085,ds003089,ds003094,ds003095,ds003096,ds003097,
                   ds003098,ds003103,ds003104,ds003114,ds003126,ds003136,
                   ds003138,ds003145,ds003146,ds003148,ds003151,ds003170,
                   ds003171,ds003176,ds003192,ds003216,ds003233,ds003242,
                   ds003338,ds003340,ds003342,ds003345,ds003346,ds003354,
                   ds003357,ds003358,ds003367,ds003381,ds003382,ds003392,
                   ds003397,ds003401,ds003404,ds003416,ds003424,ds003425,
                   ds003430,ds003433,ds003434,ds003436,ds003437,ds003438,
                   ds003439,ds003440,ds003441,ds003442,ds003443,ds003444,
                   ds003445,ds003446,ds003452,ds003453,ds003454,ds003455,
                   ds003459,ds003465,ds003466,ds003469,ds003470,ds003481,
                   ds003484,ds003487,ds003495,ds003499,ds003500,ds003505,
                   ds003507,ds003508,ds003511,ds003521,ds003524,ds003540,
                   ds003542,ds003545,ds003548,ds003550,ds003553,ds003554,
                   ds003563,ds003568,ds003569,ds003574,ds003592,ds003604,
                   ds003606,ds003612,ds003633,ds003639,ds003643,
                   ds003653,ds003659,ds003661,ds003669,ds003673,ds003684,
                   ds003688,ds003691,ds003696,ds003701,ds003707,ds003709,
                   ds003711,ds003714,ds003715,ds003716,ds003717,ds003720,
                   ds003721,ds003745,ds003752,ds003758,ds003763,ds003764,
                   ds003768,ds003770,ds003772,ds003777,ds003778,ds003782,
                   ds003787,ds003789,ds003791,ds003798,ds003799,ds003814,
                   ds003823,ds003826,ds003831,ds003834,ds003835,ds003836,
                   ds003848,ds003849,ds003851,ds003858,ds003871,ds003872,
                   ds003877,ds003892,ds003900,ds003922,ds003927,ds003949,
                   ds003950,ds003965,ds003967,ds003972,ds003974,ds003988,
                   ds003990,ds003993,ds003999,ds004006,ds004007,ds004009,
                   ds004021,ds004024,ds004038,ds004042,ds004044,ds004054,
                   ds004056,ds004065,ds004073,ds004078,ds004081,ds004086,
                   ds004091,ds004094,ds004097,ds004101,ds004102,ds004103,
                   ds004107,ds004109,ds004128,ds004129,ds004130,ds004131,
                   ds004134,ds004141,ds004142,ds004144,ds004146,ds004158,
                   ds004169,ds004173,ds004182,ds004187,ds004192,ds004194,
                   ds004196,ds004199,ds004212,ds004213,ds004217,ds004219,
                   ds004226,ds004228,ds004230,ds004259,ds004261,ds004271,
                   ds004274,ds004280,ds004283,ds004285,ds004286,ds004299,
                   ds004301,ds004302,ds004312,ds004323,ds004327,ds004331,
                   ds004332,ds004341,ds004346,ds004349,ds004359,ds004392,
                   ds004393,ds004400,ds004401,ds004406,ds004440,ds004443,
                   ds004450,ds004455,ds004456,ds004458,ds004466,ds004467,
                   ds004469,ds004470,ds004471,ds004473,ds004475,ds004478,
                   ds004482,ds004484,ds004488,ds004489,ds004493,ds004496,
                   ds004498,ds004499,ds004505,ds004513,ds004516,ds004529,
                   ds004533,ds004539,ds004542,ds004544,ds004553,ds004556,
                   ds004557,ds004560,ds004564,ds004581,ds004589,ds004590,
                   ds004592,ds004594,ds004597,ds004604,ds004605,ds004611,
                   ds004627,ds004630,ds004631,ds004636,ds004645,ds004647,
                   ds004648,ds004650,ds004654,ds004656,ds004662,ds004663,
                   ds004666,ds004670,ds004692,ds004693,ds004697,ds004698,
                   ds004711,ds004712,ds004715,ds004717,ds004718,ds004720,
                   ds004725,ds004730,ds004731,ds004733,ds004737,ds004743,
                   ds004746,ds004765,ds004775,ds004776,ds004783,ds004786,
                   ds004787,ds004791,ds004795,ds004798,ds004808,ds004814,
                   ds004815,ds004829,ds004835,ds004837,ds004848,ds004856,
                   ds004866,ds004868,ds004869,ds004873,ds004884,ds004889,
                   ds004892,ds004894,ds004909,ds004910,ds004917,ds004920,
                   ds004928,ds004934,ds004937,ds004946,ds004956,ds004957,
                   ds004958,ds004965,ds004993,ds004996,ds005003,ds005009,
                   ds005012,ds005016,ds005017,ds005025,ds005026,ds005027,
                   ds005038,ds005040,ds005047,ds005050,ds005056,ds005063,
                   ds005069,ds005072,ds005073,ds005075,ds005085,ds005088,
                   ds005096,ds005115,ds005118,ds005123,ds005125,ds005126,
                   ds005127,ds005128,ds005134,ds005139,ds005148,
                   ds005165,ds005166,ds005169,ds005191,ds005194,ds005203,
                   ds005215,ds005216,ds005226,ds005230,ds005234,ds005237,
                   ds005238,ds005239,ds005250,ds005256,ds005263,ds005264,
                   ds005265,ds005266,ds005267,ds005270,ds005279,ds005295,
                   ds005299,ds005304,ds005329,ds005341,ds005345,ds005346,
                   ds005355,ds005357,ds005360,ds005364,ds005365,ds005366,
                   ds005371,ds005374,ds005375,ds005386,ds005405,ds005412,
                   ds005413,ds005415,ds005418,ds005427,ds005449,ds005454,
                   ds005455,ds005460,ds005464,ds005468,ds005469,ds005479,
                   ds005492,ds005498,ds005503,ds005504,ds005518,ds005525,
                   ds005529,ds005530,ds005531,ds005533,ds005551,ds005559,
                   ds005571,ds005572,ds005573,ds005574,ds005576,ds005577,
                   ds005581,ds005588,ds005595,ds005596,ds005597,ds005598,
                   ds005600,ds005602,ds005603,ds005604,ds005613,ds005619,
                   ds005623,ds005625,ds005639,ds005664,ds005684,ds005699,
                   ds005700,ds005704,ds005731,ds005733,ds005737,ds005747,
                   ds005752,ds005754,ds005783,ds005795,ds005810,ds005849,
                   ds005850,ds005874,ds005875,ds005880,ds005882,ds005883,
                   ds005884,ds005891,ds005892,ds005894,ds005896,ds005899,
                   ds005901,ds005903,ds005917,ds005920,ds005947,ds005980,
                   ds006005,ds006010,ds006012,ds006033,ds006035,ds006039,
                   ds006040,ds006045,ds006067,ds006072,ds006092,ds006105,
                   ds006111,ds006128,ds006131,ds006156,ds006169,ds006179,
                   ds006181,ds006185,ds006188,ds006193,ds006206,ds006209,
                   ds006211,ds006239,ds006248,ds006265,ds006266,ds006267,
                   ds006303,ds006334,ds006391,ds006395,ds006401,ds006444,
                   ds006472}} & \multirow{4}{*}{37,949} & \multirow{4}{*}{45,377} & \multirow{4}{*}{140,389} & T1w, MP2RAGE, T2w, T2w*, T1ce,  & \multirow{4}{*} {\href{https://docs.openneuro.org/faq}{CC0}} 
                   \\ & & & & FLAIR, DWI, T1map, T2map, & 
                   \\ & & & & T2*map, GRE, ANGIO, CBF, ASL, &
                   \\ & & & & M0, minIP, SWI & \\
   SLIM~\cite{SLIM, Biswal2010} & 594 & 1,048 & 3,096 & T1w, DWI & \href{https://fcon_1000.projects.nitrc.org/indi/retro/southwestuni_qiu_index.html}{CC BY-NC}\\ 
   Tao Wu~\cite{TaoWu, Biswal2010} & 40 & 40 & 40 & T1w & \href{https://portal.conp.ca/dataset?id=projects/Taowu}{CC BY-NC-SA}\\ 
   \multirow{2}{*}{WAND~\cite{Mcnabb2025,Mcnabb2024a}} & \multirow{2}{*}{177} & \multirow{2}{*}{850} & \multirow{2}{*}{7,132} & T1w, MP2RAGE, T2w, DWI, ANGIO, & \multirow{2}{*}{\href{https://gin.g-node.org/doi/WAND/src/master/LICENSE}{CC BY 4.0}} \\
                          & & & & CBF, ASL, M0 & \\ 
   Wayne~\cite{Wayne, Biswal2010} & 426 & 610 & 610 & T1w & \href{http://fcon_1000.projects.nitrc.org/indi/retro/wayne_10.html}{CC BY-NC-SA}\\ 
   Yale Brain Mets Longitudinal~\cite{Chadha2025, Chadha2025b} & 1,430 & 11,877 & 33,800 & T1w, T2w, FLAIR, T1ce & \href{https://www.cancerimagingarchive.net/collection/yale-brain-mets-longitudinal/}{CC BY 4.0}\\ 
   Yale High Res~\cite{YaleHighRes} & 232 & 268 & 576 & T1w & \href{http://fcon_1000.projects.nitrc.org/indi/retro/yale_hires.html}{CC BY-NC-SA}\\ 
   Age ility~\cite{Karayanidis2016,age-ility} & 131 & 131 & 1,179 & T1w, DWI & \href{https://www.nitrc.org/projects/age-ility/}{CC BY-NC-SA} \\ 
\\ 

\midrule
    \textbf{FOMO260K} & \textbf{55,378} & \textbf{77,589} & \textbf{260,927} &  & CC BY-NC-SA 4.0 \\ 
\bottomrule
\end{tabular}
\caption{\textbf{Overview of the current datasets in FOMO260K}. For each dataset, we summarize the number of subjects, MRI sessions, scans, available sequence types, and license. Sequences are abbreviated as follows: T1 = T1-weighted, T2 = T2-weighted, T2* = T2*-weighted, T1ce = T1-weighted contrast-enhanced, FLAIR = Fluid-Attenuated Inversion Recovery, DWI = Diffusion-Weighted Imaging, PD = Proton Density, SWI = Susceptibility Weighted Imaging, GRE = Gradient Echo, minIP = minimum intensity projection, MP2RAGE = Magnetization-Prepared 2 Rapid Acquisition Gradient Echoes, ANGIO = Angiography, ASL = Arterial Spin Labeling, M0 = Proton Density reference image for ASL, CBF = Cerebral Blood Flow, T1map = T1 relaxation time map, T2map = T2 relaxation time map, R1map = R1 (1/T1) relaxation rate map, R2map = R2 relaxation rate map. 
}
\label{tab:fomo260kDatasets}
\end{table}

\subsection*{MRI preprocessing}

We minimally preprocessed all the MRI scans to ensure consistent brain orientation and comparable input dimensionality across datasets. To achieve this, all scans were reoriented to RAS (Right-Anterior-Superior) orientation, 4D volumes were converted to 3D, and scans with fewer than 15 slices were discarded due to insufficient anatomical coverage. 
For DWI comprising several frames corresponding to diffusion-weighted images with different gradient directions, each volume's b-value and gradient vector were used for selection. If b0 volume (b=0) or a near-b0 volume (b$\le$5) was available, it was extracted and saved separately as a reference to provide a standard non-diffusion-weighted reference. The remaining diffusion-weighted volumes were grouped into b-value shells (allowing up to 50 units of difference), and for each shell, three volumes whose gradient directions were closest to the canonical x, y, and z axes were selected (measured in terms of cosine similarity). To increase variability in DWI representations during SSL pretraining, two preprocessing strategies were applied: for half of the scans, these three volumes were averaged to produce a single 3D image representing the diffusion contrast of the shell; for the other half, a trace apparent diffusion coefficient (ADC) image was computed using the three axis-aligned volumes and the b0 reference according to:
\begin{align*}
    \text{Trace} = \text{ADC}_x + \text{ADC}_y + \text{ADC}_z, \quad 
\text{where } \text{ADC}_i = -\frac{1}{b_i} \ln \frac{S_i}{S_0}
\end{align*}
and where $S_i$ is the signal intensity of the diffusion-weighted volume along axis $i$, $S_0$ is the signal intensity of the corresponding b0 or near-b0 reference volume, and $b_i$ is the b-value corresponding to the $i$-th axis-aligned volume. To ensure numerical stability and exclude noise artifacts, ADC was computed only for voxels where $S_0 > 0$, $S_i > 0$, and $S_i \le S_0$, with the signal ratio clipped to $[10^{-10}, 1.0]$ before taking the logarithm; voxels failing these criteria were set to zero for that direction. If no b0 or near-b0 reference was available, the three axis-aligned volumes were averaged instead.
For perfusion imaging, preprocessing depended on the number of channels. When more than three channels (e.g., multiple M0 volumes) were present, all channels were averaged. For scans with two or three channels, such as ASL acquisitions, the perfusion-weighted volume was obtained by subtracting the first channel from the last. 

\subsection*{Demographic and MRI information collection}

Demographic information, including age, sex, handedness, and diagnostic group, as well as MRI acquisition details, was collected for all scans. Subject-level information was extracted from dataset-provided files or, when unavailable, from the original publications or general dataset descriptions. Each MRI session was assigned to the ``control'' groups when they corresponded to healthy individuals or patients without any major medical, neurological, or psychiatric conditions. For visualization and summary statistics, related diagnostic groups were merged into broader categories (Controls, Brain Tumor, Mental Disorders, Dementia, Neurological Disorders, Memory Complaints, and Other) using keyword-based matching. The complete group mapping algorithm is available in the publicly available data processing code repository. When information could not be determined with confidence, it was omitted. MRI acquisition metadata were obtained either from files accompanying the scans or from the original publications, including general scanner and sequence information. 

\subsection*{Dataset Composition and Demographics}

Figure~\ref{fig:dataset_overview} summarizes the overall composition of the dataset. A total of 57,224 MRI sessions have available age information at each MRI session, spanning the entire lifespan with a distribution peaking in young adulthood (mean age 34.8 years). The sex distribution is relatively balanced, with a slight predominance of females (53\%). Handedness data are available for 13,768 MRI sessions, with the majority of the subjects being right-handed (91\%). The dataset includes a broad range of subject groups. Controls account for 64\% of the MRI sessions with available diagnostic information. The remaining sessions comprise individuals with brain tumors (24\%), stroke (4\%), mental disorders (3\%), neurological disorders (1\%), and other clinical conditions (2\%). MRI data were acquired across multiple scanner manufacturers, predominantly Siemens (80\%) and Philips (10\%). Field strengths are primarily 3T (81\%) and 1.5T (15\%), with both low-field and ultra-high-field scans also represented. The dataset includes both 2D (54\%) and 3D (46\%) acquisitions. The most common sequence types are diffusion-weighted imaging (31\%), T1-weighted (31\%), and T2-weighted (12\%). Slice thickness varies across datasets, reflecting the heterogeneous acquisition protocols used across sites.

\begin{figure}[!ht]
    \centering
    \includegraphics[width=\linewidth]{figures/dataset_overview.png}
    \caption{(A) Age distribution of participants at each MRI session. (B) Distribution of subject groups at each MRI session. (C) Sex distribution at each MRI session (F = female, M = male). (D) Handedness distribution at each MRI session (R = right-handed, L = left-handed, A = ambidextrous). (E) Distribution of MRI scanner manufacturers across all scans. (F) Acquisition types showing the proportion of 2D and 3D sequences. (G) Field strength distribution. (H) Top 15 scanner models, with the MRI scanner manufacturer indicated by color coding. (I) Slice thickness distribution across the dataset (note: may include thickness of already resampled data from source datasets). (J) Top 15 MRI sequences, with modality indicated by color coding. Percentages are reported relative to the subset of data entries with available information for each variable.}
    \label{fig:dataset_overview}
\end{figure}

\subsection*{FOMO45K}

FOMO45K contains 46,149 MRI scans from 11,967 MRI sessions across 9,490 subjects, aggregated from 13 publicly available datasets. The scans in FOMO45K correspond to a subset of those included in FOMO260K, but the two releases differ in their preprocessing. They are intended to serve complementary purposes: FOMO260K provides minimally preprocessed scans that preserve the original image characteristics, whereas FOMO45K provides co-registered and either skull-stripped or defaced versions of the same scans, ready for downstream tasks that require standardized, anatomically aligned inputs (see preprocessing details below). Table~\ref{tab:fomo45kDatasets} summarizes the source datasets, including the number of subjects, sessions, and scans; MRI sequence types; preprocessing steps applied; and licenses. All included datasets are publicly available under licenses compatible with CC BY-NC-SA redistribution, and the corresponding citations provide the DOI, URL, or accession identifier required to retrieve each original dataset.

\begin{table}[!ht]
\centering
\scriptsize
\setlength{\tabcolsep}{6pt}

\begin{tabular}{l r r r l l l}
\toprule
\makecell{\textbf{Source dataset}} &
\makecell{\textbf{Subjects}} &
\makecell{\textbf{MRI sessions}} &
\makecell{\textbf{MRI scans}} &
\makecell{\textbf{MRI sequences}} &
\makecell{\textbf{Skull-stripped / Defaced}} &
\makecell{\textbf{License}} \\
\midrule
    OpenNeuro - SOOP~\cite{ds004889} & 1,715 & 1,715 & 6,508 & T1, FLAIR, DWI & Defaced & \href{https://docs.openneuro.org/faq}{CC0}\\ 
    BraTS-GEN~\cite{BraTS-GEN, Labella2023, Adewole2023, baid2021, Moawad2024, Kazerooni2024} & 2,251 & 2,251 & 9,004 & T1, T2, FLAIR, T1ce & Skull-stripped & \href{https://www.synapse.org/Synapse:syn53708249/wiki/627759}{CC BY-NC} \\ 
    MSD Brain Tumor~\cite{Simpson2019, Menze2014, Bakas2017, Bakas2018, MSD} & 484 & 484 & 1,936 & T1, T2, FLAIR, T1ce & Skull-stripped & \href{http://medicaldecathlon.com/}{CC BY-SA 4.0} \\ 
    IXI~\cite{IXI} & 584 & 584 & 2,530 & T1, T2, PD, DWI & Skull-stripped & \href{https://brain-development.org/ixi-dataset/}{CC BY-SA 3.0}\\ 
    OpenNeuro - NIMH~\cite{ds005752} & 249 & 252 & 2,089 & T1, T2, T2*, DWI & Skull-stripped  & \href{https://docs.openneuro.org/faq}{CC0}\\ 
    OpenNeuro - DLBS~\cite{ds004856} & 464 & 957 & 3,845 & T1, T2, DWI & Skull-stripped  & \href{https://docs.openneuro.org/faq}{CC0}\\ 
    OpenNeuro - IDEAS~\cite{ds005602} & 542 & 542 & 1,035 & T1, FLAIR & Defaced  & \href{https://docs.openneuro.org/faq}{CC0}\\ 
    OpenNeuro - ARC~\cite{ds004884} & 230 & 741 & 2,029 & T1, T2, DWI & Defaced  & \href{https://docs.openneuro.org/faq}{CC0}\\ 
    OpenNeuro - MBSR~\cite{ds005016} & 147 & 348 & 1,023 & T1, DWI & Defaced  & \href{https://docs.openneuro.org/faq}{CC0}\\ 
    OpenNeuro - UCLA~\cite{ds000030} & 265 & 265 & 789 & T1, DWI & Defaced  & \href{https://docs.openneuro.org/faq}{CC0}\\ 
    OpenNeuro - QTAB~\cite{ds004146} & 417 & 721 & 6,760 & minIP, GRE, SWI & Defaced  & \href{https://docs.openneuro.org/faq}{CC0}\\ 
    NKI~\cite{Tobe2022, NKI} & 1,316 & 2,281 & 4,502 & T1, T2, DWI & Defaced & \href{https://rocklandsample.org/for-researchers/step-2-data-usage-agreement}{Open Access}\\ 
    OpenNeuro - AOMIC ID1000~\cite{ds003097} & 826 & 826 & 4,099 & T1, DWI & Defaced  & \href{https://docs.openneuro.org/faq}{CC0}\\
\midrule
\textbf{FOMO45K} & \textbf{9,490} & \textbf{11,967} & \textbf{46,149} &  &  & CC BY-NC-SA 4.0 \\
\bottomrule
\end{tabular}
\caption{\textbf{Overview of the datasets in FOMO45K}. For each dataset, we summarize the number of subjects, MRI sessions, scans, available sequence types, whether the images were skull-stripped or defaced, and license. See Methods for full preprocessing details. Sequences are abbreviated as follows: T1 = T1-weighted, T2 = T2-weighted, T2* = T2*-weighted, T1ce = T1-weighted contrast-enhanced, FLAIR = Fluid-Attenuated Inversion Recovery, DWI = Diffusion-Weighted Imaging, PD = Proton Density, SWI = Susceptibility Weighted Imaging, GRE = Gradient Echo, minIP = minimum intensity projection.}
\label{tab:fomo45kDatasets}
\end{table}

MRI preprocessing for FOMO45K comprised three main steps: reorienting images to RAS orientation (as in FOMO260K), affine co-registration, and skull-stripping. All scans were first reoriented to RAS and affinely co-registered using the \texttt{mri\_coreg} command from FreeSurfer 7.4.1~\cite{Fischl2012}, with default parameters. Within each MRI session, scans were aligned to the image with the highest spatial resolution to preserve most anatomical detail.

For DWI scans in 4D format, if a b0 volume was available, it was extracted and saved separately. For the b1000 shell, the volumes were processed as in FOMO260K and averaged into a single representative 3D image. In contrast to FOMO260K, no trace ADC computation or additional b-value shell grouping was performed.

Skull-stripping was performed using SynthSeg~\cite{Billot2023} (FreeSurfer 7.4.1), which outputs segmentation masks of brain structures. These masks were used to define the brain extraction region. Skull-stripping was only applied when the images were not already defaced or skull-stripped by the dataset provider.

\subsection*{Ethics statement}

All data included in FOMO260K and FOMO45K were obtained from publicly available sources under open licenses; the original license of each constituent dataset is listed in Tables~\ref{tab:fomo260kDatasets} and \ref{tab:fomo45kDatasets}. Each constituent dataset was collected under protocols approved by the respective institutional review boards of the original studies. No new human subjects research was conducted for this data aggregation project.

\section*{Data Record}

The FOMO260K dataset is publicly available through \href{https://huggingface.co/datasets/FOMO-MRI/FOMO260K}{huggingface.co/datasets/FOMO-MRI/FOMO260K}, while FOMO45K is publicly available at \href{https://huggingface.co/datasets/FOMO-MRI/FOMO45K}{huggingface.co/datasets/FOMO-MRI/FOMO45K}. 
All MRI scans are stored in NIfTI-compressed format and organized using a modified version of the Brain Imaging Data Structure (BIDS) format~\cite{Gorgolewski2016}. For most modalities, the directory structure and naming closely follow BIDS conventions. However, several deviations were introduced to ensure scalability and consistency across FOMO260K. First, acquisition and scanner metadata are not stored as individual JSON files per scan; instead, MRI acquisition information is centralized in a single tabular file (\texttt{mri\_info.tsv}). Second, when sequence information was unavailable or ambiguous, scan names were kept close to their original labels or generically named \texttt{scan}. Third, diffusion scans do not follow the BIDS DWI specification: they are provided as derived 3D volumes without accompanying \texttt{bval} and \texttt{bvec} files, with the b-value group and whether a trace ADC representation (\texttt{\_trace}) was used explicitly encoded in the filename. Finally, subject-level information and scan provenance are provided via tabular files: \texttt{participants.tsv} contains demographic information and group assignments, while \texttt{mapping.tsv} records the correspondence between each scan and its original source dataset. Full details are documented in the accompanying code repository.

\section*{Technical Validation}

\subsection*{Empirical Validation} 
To validate the efficacy of the FOMO260K dataset, we pretrain a model on FOMO260K using a self-supervised masked autoencoder objective, in which the model learns general-purpose representations by reconstructing masked portions of the input without relying on any task-specific labels. We then show that the pretrained model achieves better downstream performance than a model trained from randomly initialized weights (referred to as ``scratch'' in the remainder of the paper).
We evaluate the result of pretraining in the challenging few-shot setting, where the model must generalize with limited supervision, making any improvement a direct indicator that the pretrained features transfer effectively to new tasks. 

\subsubsection*{Pretraining configuration}
We train a 100M-parameter ResEnc\cite{isensee_2024} model using AMAES (Augmented
Masked Auto Encoder for 3D Segmentation)\cite{munk2024amaes}, a variant of the popular masked-autoencoder pretraining strategy optimized for representation learning on 3D MRI data\cite{he2022}, using a masking ratio of 60\% and a masking unit size of $8\times8\times8$ voxels. Training used eight H100 GPUs for 44 hours with a global batch size of $64$ for 187,500 steps. Each scan was processed using a patch size of $160\times160\times160$. We optimize with AdamW~\cite{Loshchilov2017} at a base learning rate of $1\times10^{-4}$ under a cosine decay schedule, with 2\% linear warmup and gradient accumulation of two steps.

\subsubsection*{Finetuning configuration}
We evaluate models on a diverse set of segmentation tasks using the ISLES22~\cite{Hernandez2022} ($n=250$), ATLAS~\cite{Liew2022} ($n=655$),  SBM3~\cite{SBM, Grovik2019} ($n=105$), WMH\cite{wmh} ($n=170$), and Cerebrum-7T \cite{Svanera2021} ($n=142$) datasets. All datasets are used in a few-shot setting, using 20 labeled examples per training split. Models are trained for 37,500 steps with AdamW, a max learning rate of $1\times10^{-3}$. Only the encoder weights are transferred. To adapt the decoder to the encoder, we first freeze the encoder for 2500 steps, training only the decoder with a linearly increasing learning rate. The encoder and decoder are then linearly warmed up for 2,500 steps before the entire network is trained with a cosine decay schedule. We use a batch size of two. Scratch baselines follow an nnU-Net-inspired configuration~\cite{Isensee2021}, employing stochastic gradient descent with a base learning rate of $1\times10^{-2}$ using a cosine decay schedule for 37,500 steps.

\subsubsection*{Validation results}
Results are provided in Table~\ref{tab:results}. We report the average Dice score over five folds along with the standard error of the mean. If the dataset contained multiple classes, we report the average of the foreground classes. The pretrained model consistently outperforms the scratch baseline across all five benchmark datasets, with the differences statistically significant at the $p<0.05$ level (two-sided paired t-test on per-case Dice differences, Holm-corrected~\cite{Holm1979} across datasets). The largest gains are observed on SBM3 ($+1.45$ Dice), WMH ($1.29$ Dice), and ISLES22 ($+1.12$ Dice), with smaller but still consistent gains on ATLAS, and Cerebrum-7T.

\section*{Usage Notes}

FOMO260K~\cite{FOMO260K} and FOMO45K~\cite{FOMO45K} are distributed under the CC BY-NC-SA 4.0 license. The original licensing terms of each constituent dataset are listed in Table~\ref{tab:fomo260kDatasets} for FOMO260K and Table~\ref{tab:fomo45kDatasets} for FOMO45K. To ensure proper attribution and recognition of the source datasets, all users must cite the following papers for FOMO45K and FOMO260K: BraTS-GEN~\cite{Labella2023, Adewole2023, baid2021, Moawad2024, Kazerooni2024}, MSD Brain Tumor~\cite{Simpson2019, Menze2014, Bakas2017, Bakas2018}, IXI~\cite{IXI}, NKI~\cite{Tobe2022}. 

Additionally, the following papers must be cited for FOMO260K: ClevelandCCF~\cite{ClevelandCCF}, Nigerian Clinical~\cite{Wogu2025}, CUNMET~\cite{CUNMET}, ACPI~\cite{ACPI}, ADHD\_200~\cite{ADHD200}, AHEAD~\cite{Alkemade2020}, ATAG~\cite{Forstmann2014}, Adolescent Brain Development~\cite{Geeraert2020}, CFMM-7T~\cite{Haast2021}, CHBMP~\cite{Valdes2021}, Calgary Preschool~\cite{Reynolds2020}, CoRR~\cite{CoRR}, HBN-SSI~\cite{HBN-SSI, Biswal2010}, HBN~\cite{Alexander2017}, Beijing Enhanced~\cite{BeijingEnhanced, Biswal2010}, Infant Development Brain~\cite{Akinci2023}, M4Raw~\cite{Lyu2023}, SLIM~\cite{SLIM, Biswal2010}, Tao Wu~\cite{TaoWu, Biswal2010}, WAND~\cite{Mcnabb2025}, Wayne~\cite{Wayne, Biswal2010}, Yale Brain Mets Longitudinal~\cite{Chadha2025, Chadha2025b}, Yale High Res~\cite{YaleHighRes}, and Age ility~\cite{Karayanidis2016}.

\begin{table}[!t]
\centering
\scriptsize
\begin{tabular}{l c c c c c}
\toprule
Method &
ATLAS &
ISLES22 &
SBM3 & 
WMH & 
Cerebrum-7T \\
\midrule
Scratch  & $41.49 \pm 0.72$ & $72.86 \pm 0.89$ & $58.23 \pm 1.44$ & $72.75 \pm 0.58$ & $90.54 \pm 0.12$ \\
AMAES on FOMO260K & $\mathbf{42.43 \pm 1.36} $ & $\mathbf{73.98 \pm 0.64}$ & $\mathbf{59.68 \pm 1.03}$ & $\mathbf{74.04 \pm 0.51}$ & $\mathbf{90.80 \pm 0.06}$ \\
\bottomrule
\end{tabular}
\caption{Dice performance (mean $\pm$ standard error of the mean) across five-fold cross-validation. Pretraining on FOMO260K yields consistently higher accuracy than training from scratch. \textbf{Bold} values indicate $p < 0.05$ for a two-sided paired t-test on per-case Dice differences, with Holm correction~\cite{Holm1979} across datasets.}
\label{tab:results}
\end{table}

\section*{Data Availability}

FOMO260K and FOMO45K are publicly available on Hugging Face at 
\url{https://doi.org/10.57967/hf/8670} and \url{https://doi.org/10.57967/hf/8669}, respectively, 
under the CC BY-NC-SA 4.0 license.

\section*{Code Availability}

The preprocessing scripts for FOMO260K are publicly available at \url{https://github.com/Sllambias/asparagus_preprocessing}. The code used for pretraining the models is available at \url{https://github.com/Sllambias/asparagus}. The preprocessing scripts for FOMO45K, together with additional code required to reproduce all analyses, tables, and figures reported in this manuscript, are available at \url{https://github.com/FGA-DIKU/fomo_mri_datasets}. Pretrained model weights are publicly released at \url{https://huggingface.co/FOMO-MRI/AMAES_resenc_b_fomo260k}.

\bibliography{references}

\section*{Author Contributions}
S.C. and A.M. contributed equally to this work and were responsible for conceptualization, methodology, data curation, software development, formal analysis, validation, visualization, and writing of the original draft.
S.N.L. contributed to methodology, software development, data curation, and manuscript review and editing.
J.A. contributed to methodology, software development, formal analysis, and manuscript review and editing.
J.M. contributed to software development, data curation, and manuscript review and editing.
P.R.G. contributed to software development and manuscript review and editing.
V.N., C.H.K., P.L., M.M.G., M.B., and M.E.B. contributed to manuscript review and editing.
J.E.I. contributed to supervision, resources, and manuscript review and editing.
M.N. contributed to supervision and manuscript review and editing.
All authors reviewed and approved the final manuscript.

\section*{Competing Interests}

M.N. holds shares in Cerebriu.

\section*{Acknowledgements}

\begin{itemize}
    \item BraTS-GEN: Data used in this publication were obtained as part of the Challenge project through Synapse ID (syn53708249)
    \item Beijing Enhanced: Financial support for the data used in this project was provided by a grant from the National Natural Science Foundation of China: 30770594 and a grant from the National High Technology Program of China (863): 2008AA02Z405.
\end{itemize}

\section*{Funding}

This work has been supported by the Danish Data Science Academy, which is funded by the Novo Nordisk Foundation (grant number NNF21SA0069429) and Villum Fonden (grant number 40516), Pioneer Centre for AI, Danish National Research Foundation, grant number P1, the Lundbeck Foundation (grant number R449-2023-1512), the Novo Nordisk Foundation (grant number 0104988 for the Gefion AI Supercomputer), and the National Institute of Health (grant number 1R01AG070988, 1RF1AG080371, 1RF1MH123195,  1UM1MH130981, 1R21NS138995, and 1R01EB031114).

\end{document}